# Dreams versus Reality: Plenary Debate Session on Quantum Computing

6:00pm, Wednesday, 4th June 2003, La Fonda Hotel, Santa Fe, USA
Part of: SPIE's First International Symposium on Fluctuations and Noise (FaN'03)
http://spie.org/conferences/programs/03/fn/

## The Panel – Dramatis Personae

*Chair/Moderator:* **Charles R. Doering**, Univ. of Michigan (USA);
Editor of *Physics Letters A*

*Pro Team:*
**Carlton M. Caves**, Univ. of New Mexico (USA);
**Daniel Lidar**, Univ. of Toronto (Canada); Editor of *Quantum Information Processing;*
**Howard Brandt**, Army Research Lab. (USA);
**Alex Hamilton**, Univ. of New South Wales (Australia).

*Con Team:*
**David Ferry**, Arizona State University (USA); Editor of *Journal of Computational Electronics; Journal of Applied Physics/Applied Physics Letters; Solid State Electronics; Superlattices and Microstructures*
**Julio Gea-Banacloche**, Univ. of Arkansas (USA); Editor of *Physical Review A;*
**Sergey Bezrukov**, National Institutes of Health (USA); Editor of *Fluctuation Noise Lett.;*
**Laszlo Kish**, Texas A&M (USA); Editor-in-Chief of *Fluctuation Noise Lett.*

*Transcript Editor:*
**Derek Abbott**, The University of Adelaide (Australia); Editor of *Fluctuation Noise Lett; Smart Materials and Structures.*

**Disclaimer:** The views expressed by all the participants were for the purpose of lively debate and do not necessarily express their actual views.

**Transcript conventions:**
- Square brackets […] containing a short phrase indicate that these words were not actually spoken, but were editorial insertions for clarity.
- Square brackets […] containing a long section indicate that the recording was unclear and the speaker has tried to faithfully reconstruct what was actually said.
- [*sic*] indicates the transcript has been faithful to what was spoken, even though grammatically incorrect.
- Angular brackets <…> indicate editorial comments by the Transcript Editor.
- Where acoustic emphasis was deemed to occur in the recording, the transcript reflects this with italics.

# The Debate

**Chair (Charlie Doering):** [Welcome everybody to the Plenary Debate on quantum computing "Dream or Reality." We are going to start with the Pro team and then the Con team. Each speaker will strictly have 3 minutes. Before we start, I would like to remind everybody about the dangers of trying to make future predictions about computers with the following quote:]

"I think there is a world market for maybe five computers."
— Thomas Watson, Chairman of IBM, 1943.

**<Audience laughter>**

**Chair (Charlie Doering):** [OK, now let's move straight to our first panellist. Carl.]

**First Pro Panelist (Carl Caves):** [I'm going to declare that the subject of this debate is the question: "Is it possible to build a quantum computer?" With this question in hand, we still have to define our terms. What does "possible" mean? It could mean, "Are quantum computers allowed by physical law?" Since we think they are, and since small numbers of qubits have been demonstrated, I'm going to define "possible" to mean, "Can it be done in $n$ years?" And then we have the further question of the value of $n$. Does $n = 1,000$, $n = 100$, $n = 30$ or $n = 10$? Finally, we need to define what we mean by a "quantum computer." Do we mean a rudimentary, but scalable device that can, say, factor 15? Do we mean a useful quantum simulator? Or do we mean a scalable, general-purpose quantum computer (e.g., one that factors interestingly large numbers)?

Before proceeding, I will issue a warning about physicists' estimates of the time needed to accomplish some task:

$n = 1$ year: This is a reliable guess, but it will probably take 2 to 5 years.

$n = 10$ years: I have a clue how to proceed, but this is a guess that I'm hoping the funders will forget before the 10 years are out.

$n > 30$ years: I don't have a clue, but someone put a gun to my head and made me guess.]

**<Audience laughter>**

**First Pro Panelist (Carl Caves):** [So $n$ is going to have a different answer depending on what we mean. Here are my estimates for the three cases:

- Rudimentary, but scalable device that can, say, factor 15? — 10-15 years
  Motivation: High

- Useful quantum simulator? — 20-30 years
  Motivation: Medium

- General-purpose quantum computer? — 50 years
  Motivation: Need more algorithms

The really important question for discussion is not whether we can build a quantum computer, but rather, "Is quantum information science a worthwhile interdisciplinary research field?" Yes! It puts physical law at the core of information-processing questions. It prompts us to ask what can be accomplished in a quantum-mechanical world that can't be accomplished in a classical world. And it prompts us to investigate how to make quantum systems do what we want instead of what comes naturally.]

**Chair (Charlie Doering):** [OK, good timing. Next, Daniel.]

**Second Pro Panelist (Daniel Lidar):** I think it was in '95, <a paper in *Physics Today* by Haroche and Raimond, > and the title of the paper was "Quantum computing: dream or nightmare," – so we're making progress by making this [debate] "dream or reality."

**<Audience laughter>**

**Second Pro Panelist (Daniel Lidar):** I believe that there's no question that quantum computers will be built and my reasoning for this is very simple. There is simply no law of nature, which prevents a quantum computer from being built and there are some damn good reasons for building one. Now, why is it that there is no law of nature? Well, in that 'dream or nightmare' paper and as well as some other papers by Bill Unruh and Landauer around the same time, it was argued that decoherence was going to kill quantum computers and therefore there was no chance. And there was a quick reaction to that, which astonished a lot of people because it seemed to somehow violate the second law of thermodynamics; and this was the discovery of quantum error correcting codes.

So while it was believed naïvely that quantum computers would never work because of decoherence, quantum error correcting theory shows that this belief was false and that in fact it is possible to overcome the decoherence problem, at least in principle. And this theory has been refined to the level where we now know that there exists a threshold, which is measured in terms of a number that's rather small – about $10^{-4}$ or so. It's basically something like the ratio between the decoherence time and the time it takes to apply an elementary logic gate. So you have to be able to squeeze in $10^4$ logic gates within a unit of decoherence time.

If you can do that, we know from the theory of quantum error correction that there is nothing in principle preventing quantum computers from being built. They will be robust; they will be able to resist decoherence absolutely. So this disproves the old skepticism and I believe it reduces the problem of constructing a quantum computer to a very interesting one, but it has basically now become a problem of finding the optimal system and fine-tuning the ways that we're going to implement quantum error correction, quantum logic gates and measurements.

**Chair (Charlie Doering):** Excellent. Thank you. Twenty seconds to spare.

**<Audience applause>**

**Chair (Charlie Doering):** Now can we have Howard?

**Third Pro Panelist (Howard Brandt):** Yes.

**Chair (Charlie Doering):** OK, go!

**Third Pro Panelist (Howard Brandt)**: Quantum information processors are certainly viable: quantum crypto systems, operational quantum crypto systems, have already been demonstrated. That's small-scale quantum information processing. Quantum teleportation has been demonstrated. Some of the basic ingredients of quantum repeaters have been demonstrated. Quantum copiers are certainly feasible. Grover's algorithm has had a proof-of-principle – Grover's Algorithm has been demonstrated for a database of 8 entries, doing it faster than a classical computer could have. So the algorithm has been proof-of-principle demonstrated. Shor's algorithm has been proof-of-principle demonstrated in factoring the number 15 faster than a classical computer could. Quantum error correction has also been demonstrated on a small scale – proof-of-principle. As Lidar has pointed out, the laws of physics do not prohibit even large-scale quantum computers capable of exercising Shor's algorithm, or Grover's algorithm to search a large database.

There's potentially a big pay-off in solving problems *not possible* to solve classically, and breaking unbreakable codes. Speaking for the viability of quantum information processors is the worldwide effort including many elements, many disciplines, special sections of our most prestigious journals on quantum information, and a number of entirely new journals on this subject. The real feasibility of developing a robust large-scale quantum computer by any of the current approaches remains in question. It will likely take a lot of time.

**Chair (Charlie Doering)**: 30 seconds.

**Third Pro Panelist (Howard Brandt)**: Luv Grover has warned us against erring on the side of pessimism. Witness the pessimism at the end of the ENIAC in 1949 in terms of projected size and number of vacuum tubes.

**Chair (Charlie Doering)**: One second.

**Third Pro Panelist (Howard Brandt)**: My time is up, is it? All right, well…the Army Research Lab is tasked with assessing the viability of quantum information science and technology for possible transition to the Army. Quantum crypto systems are ready and that's being pursued – the other systems are not ready but they will be.

**Chair (Charlie Doering)**: OK, thank you very much.

**<Audience applause>**

**Chair (Charlie Doering):** OK, I'm pretty excited now! Alex.

**Fourth Pro Panelist (Alex Hamilton)**: OK, so to follow Carl's theme; the first thing we've got to do is look at the question: Is it a dream or reality? And the answer is we best not follow that path, actually, because it is an entanglement of both. The dream is really to – as system engineers – to understand nature and to try to control nature. What's the simplest quantum mechanical thing we can understand? [It is] the quantum two-level system. What *could* be simpler? Let's get one and control it. That will be a beautiful thing to do. Understanding even what quantum mechanics means at the most fundamental level – this is all part of the dream. Well, what do we mean by doing a quantum measurement? We teach our high school and undergraduate students that you have a quantum system, you come along, do a measurement and that collapses the wave

function – but we're not really sure how it collapses the wave function – that's never really discussed. It just comes in, it collapses and you get a 1 or a 0. The cat is either alive or dead. So, we're having to think very hard about what a quantum measurement means. This seemingly esoteric and irrelevant question now has a very real physical meaning in terms of doing a measurement on a quantum bit. And then, how do you couple these two-level systems? What does it mean to entangle them? Do you actually need entanglement for quantum computing? So these are, physically, very important questions to answer.

The other breakthrough is that it does bring together people from all sorts of different disciplines. In the solid-state area there are people from superconductivity, semiconductors, surface science, other variant schools of physics, all talking about the same thing and for the first time, in a long time, speaking the same language. So there are 3 level quantum systems, decoherence, T1, T2. In liquid-state NMR, same thing is happening. Every phase of matter is being represented: solid-state, through liquid, through gas, even the others: Bose-Einstein condensates, fractional quantum Hall liquids. We're all coming together and talking the same language, so that's the dream, who knows? The reality – can it be done? Well, there's good evidence that we can make one-qubit systems. There's evidence we can couple $n$ small numbers of qubits. So on a very small scale, yes, it looks like it can be done. Can it be scaled to a usefully large quantum computer? That really is a very difficult question to answer. I would say that it is perhaps too early to say because it's a big engineering problem but there's no law of physics that says that it simply cannot be done. And, again, if we look at the history of electronics the first vacuum valves were in operation in the early twentieth century but it wasn't until about the 1960s that it was possible to really build a useful [classical] computer.

**Chair (Charlie Doering)**: Twenty seconds.

**Fourth Pro Panelist (Alex Hamilton)**: Will it actually be useful? Will I be able to go down to Walmart and buy one for my grandmother for Christmas?

**<Audience laughter>**

**Fourth Pro Panelist (Alex Hamilton)**: This is a question that one of my students asked. Maybe it will never be one per household, but perhaps we don't need it to. Supercomputers – my grandmother doesn't have a supercomputer at home – she's quite happy, in fact she doesn't have a computer at home. And perhaps you can say, 'Look, it doesn't matter. It's just never going to work.' How do we know? If you look at the history of computers, people said that it would never work. They were weighing no more than 1.5 tonnes and they'll consume no more than the power of a small city; and look what we've got today, so I think it *is* possible and it's just… you better go and see what happens.

 **Chair (Charlie Doering):** OK, thank you very much.

**<Audience applause>**

**Chair (Charlie Doering):** OK. I'm glad you addressed the issue of how much they're going to weigh. That's certainly something on a lot of people's minds. At least a while ago…

"Computers in the future may weigh no more than 1.5 tonnes." – *Popular Mechanics,* forecasting the relentless march of science, 1949.

**<Audience laughter>**

 **Chair (Charlie Doering):** Now we're going to go over to the Con side, which I believe is some kind of Republican view of …

**<Hysterical audience laughter>**

**Chair (Charlie Doering):** We'll start off with David Ferry – please.

**First Con Panelist (David Ferry):** Just a simple, little 2-level system. It would be the easiest thing in the world to make, all right? We've had *twenty* years working on quantum computers, more than two decades in fact and we haven't got it going yet. The problem is that in those two decades – more than two decades – they've only got two algorithms. Although I heard in a rumour, today, that a third algorithm may have been followed up. Without having the architecture and an algorithm making the system work, you can't make a system. So you really have to have more than just a device – the world is littered with devices and it takes more than just a theory. When I was young, last century or just before …

**<Audience laughter>**

**First Con Panelist (David Ferry):** [Many years ago,] one of the first conferences I went to was on superconductivity, and there was a god of superconductivity who made the statement that *all* the theory in the world, integrated over time had not raised the transition temperature one milliKelvin. So it takes more than just ideas about where science goes – you have to have practical working examples from the laboratory. You have to see the results. If you haven't seen it yet, it's quite a difficult problem. We've been arguing about [quantum] measurement since the 1927 Solvay Conference, and even the idea of wave-function collapse depends upon your view of quantum mechanics. [Bob] Griffiths doesn't believe in wave-function collapse. So you have to be careful now about your interpretation. This makes it a very difficult problem both intellectually and practically, but it's a dream with a shift in emphasis over there by Caves and he probably should work on quantum information. **<Chair taps to indicate time. >** OK. Great.

**<Audience applause>**

**Chair (Charlie Doering):** Excellent, excellent. The economy of your presentation was perfect. Julio?

**Second Con Panelist (Julio Gea-Banacloche):** Alright, well …I …um …I'm surprised, actually, that I'm sitting here on the Con side…

**<Audience laughter>**

**Second Con Panelist (Julio Gea-Banacloche)**: …. because I just realised that I'm actually more optimistic than Carl is.

**<Audience laughter>**

**Second Con Panelist (Julio Gea-Banacloche)**: I would like to mention, nonetheless, that the reason I'm here, I think, is because I understood the question to mean the last of these options, that is to say, the general purpose, huge, big, million physical qubit factoring machine of strategic importance and so on. And that actually …personally, I don't think that we will ever see that, for the reason that it's basically—even though we may call it a universal quantum computer, and that seems to confuse some people—we really don't mean that this is a computer that will replace current computers in any sense. We're not building this so that we can run Microsoft Office on it.

**<Audience laughter>**

**Second Con Panelist (Julio Gea-Banacloche)**: In fact, there is no reason to build anything to run Microsoft Office on it [*sic*].

**<Audience laughter>**

**Second Con Panelist (Julio Gea-Banacloche)**: …but this is obviously going to be – even if it is built – it's going to be a special purpose machine and in fact, as Carl also has pointed out, so far we have only one reason to build it... and that is to break certain encryption systems, which are currently very popular. But the thing is that this device is not going to be built, if at all, for some 20 years, 30 years or something like that. And I find it very hard to believe that in 20-30 years people are still going to be relying for the encryption of sensitive data on the same encryption algorithm that today a quantum computer can break. Given that, my personal prediction is that this idea is going to go basically the way of some other technologies that looked very promising at one time, but they turned out to be so extremely challenging that they failed to deliver on their promise in a timely fashion and they simply fell by the wayside – mostly we found ways around them, and that's basically what I think is going to happen with quantum computers; I mean the large scale quantum computers. Like Carl and like everybody else, I think that this is very valuable scientific research and having a small, say 100 qubit, quantum simulator in 10 or 15 years will be a big accomplishment and not out of the realm of possibility.

**Chair (Charlie Doering)**: Thank you.

**<Audience applause>**

**Chair (Charlie Doering)**: Sergey.

**Third Con Panelist (Sergey Bezrukov)**: The organizers have asked me to say something about quantum computing and biology. <The possible implication being that if nature hasn't somehow made use of quantum computing, itself, then there probably isn't much hope for it. > This something is a very short message, which is 'there is no place for quantum computing in our brain.' The main function [of the brain] is based on nerve pulse propagation and this process has been studied in great detail. What I mean is that most of you in this audience do not have any idea of how many people [are working on these problems] and how much effort is put into this investigation. It is well understood that this [pulse propagation] is a dissipative macroscopic process. The next in line is synaptic transduction. This is how nerve cells talk to teach other. Again, this is a macroscopic dissipative process, which is understood right down to molecular detail. Next, there are – and these are necessary for 'computation' in our brain – short-term and long-term memory. Well, these things are not as well studied as the previous two, but one can say that the short-term memory is related to the short-term changes in the chemical composition of interacting cells. For example, if I say something to you right now, you are able to recall it within time intervals of several seconds, because of the transient chemical changes in the right places. And, finally, our long-term memory is definitely related to, again, macroscopic dissipative processes leading to structural changes in the brain. This is all.

**Chair (Charlie Doering):** Thank you.

**<Audience applause>**

**Chair (Charlie Doering):** OK, everybody take a deep breath because – next – Laszlo is going to give us his view on the subject.

**Fourth Con Panelist (Laszlo Kish):** I don't have much to say… yes, it's really marvellous that the quantum field has found new effects. This is really great. My problem is with – just like Julio – general-purpose quantum computing, it seems, is like analog computing: we have to build a system that is special purpose. The error space is analog. What we have to see is that quantum parallelism is a consequence of Hilbert space. But classical systems also can inhabit Hilbert space. So to save time, you can also try classical systems. When we compare classical and quantum computing, it is very important to use the same temperature and the same speed <clock frequency> and then compare a classical hardware version with a quantum version, with the same number of elements, and ask what is the power dissipation. Another question is where are the general-purpose quantum algorithms? It is important to note that a classical Hilbert Space computer is already working in Japan! A 15-qubit classical quantum computer was built by Fujishima – we saw this in the talks. Thanks.

**Chair (Charlie Doering):** OK, thank you very much.

**<Audience applause>**

**Chair (Charlie Doering):** What we're going to do now is… I'm going to show you something to the Con side here: the Pro side is very busy taking notes while you were all speaking. So, now, I'd like to have a reality check every once in a while about telling the future of computer science:

"There is no reason anyone would want a computer in their home."
– Ken Olson, president, chairman and founder of Digital Equipment Corp., 1977.

**<Audience laughter>**

**Chair (Charlie Doering):** That's right, that's why DEC doesn't exist anymore, OK!

**<Audience laughter>**

**Chair (Charlie Doering):** Now what we're going to do is we're going to go back. Quickly, one minute, each person, same direction; any comments they want to make, any ridiculing they want to do. Then we're going to open it up to a free discussion to take comments and questions from the audience and so on OK so, we'll start off right now with Carl.

**First Pro Panelist (Carl Caves):** I'm going to try and make four quick points. First, it is good to know that the editor of *PRA* for quantum information views research in this field as useful.

**First Pro Panelist (Carl Caves):** I think we might learn that… this is in response to some comments by Dave Ferry… I think we might learn some things about how to interpret quantum mechanics by thinking in terms of quantum information processing. I think quantum mechanics is partly information theory, partly physical theory, and we've never understood exactly how these two go together. We might learn something in this regard, but I don't think we have to know anything about the interpretation of quantum mechanics to know how physicists will interpret and make predictions for what a computer will do.

I guess I want to get to my third point: I agree that we need more algorithms. Let me say that the only reason I'm optimistic about that is – because I don't know much about that and I don't think there's anyone here in this room who's a real expert on that – Umesh Vazirani told me that after probabilistic algorithms came out, it took people 15 years before they realised what could be done with probabilistic algorithms. Maybe something like that will happen with quantum algorithms.

**Chair (Charlie Doering)**: OK, thank you. No applause for this [round]. Too much time, too much time. Daniel.

**Second Pro Panelist (Daniel Lidar):** OK, let me take these guys on one by one.

**Second Pro Panelist (Daniel Lidar):** David Ferry says that "20 years and we have no qubits yet, no [new] algorithms, no practical devices" – but he neglects the amazing results in trapped ions, 4-qubit entanglement, Josephson qubits have already shown entanglement, and in quantum dots single qubit operations have already been performed. [This] all happened in the last three or four years – no reason that it won't continue. Julio says "only reason is to break crypto" but he forgets that quantum computers will be to simulate quantum mechanics exponentially faster than we can do on classical devices. Now, Sergey: "no role for quantum computers in the brain" – I agree.

**Second Pro Panelist (Daniel Lidar):** Laszlo: "quantum computers are like analog systems, that are special purpose," well, they are not analog. Actually they are digital. That is a subtle point. "Classical computers can be described in Hilbert space." Yes, but there's no entanglement, no tensor product structure. The whole speed up issue just breaks down for classical computers, even if you use Hilbert spaces.

**Chair (Charlie Doering)**: Perfect. OK. Howard.

**Third Pro Panelist (Howard Brandt):** I agree with Dan that David is not up to date. I'm not surprised, because when I looked at his paper, he speaks of entanglement as being a hidden variable… enough for David.

**First Con Panelist (David Ferry):** <clutches his chest as if he's been shot>

**Third Pro Panelist (Howard Brandt):** Julio, well, I think that you have to realise that a universal quantum computer is a mathematical artifice, as was the Turing machine. It's an idealisation – something that will be approached – it does not deal with decoherence, it doesn't deal properly with a halt bit. There are certain operations and certain unitary transformations that are suspect. However, related to the universal quantum computer – we now have a generalised Church-Turing thesis.....

**Chair (Charlie Doering)**: 10 seconds.

**Third Pro Panelist (Howard Brandt):** The original Church-Turing thesis is not true because of quantum computers. Also I heard that factoring might not be that important. But Grover's search will, and there will be other NP incomplete algorithms (such as the travelling salesman problem) that may happen. And Laszlo, sure you can use Hilbert space for some classical systems, but that's an *entirely* different ballgame than in quantum mechanics.

**Chair (Charlie Doering)**: OK, right, let's move on. You'll get another chance.

**Fourth Pro Panelist (Alex Hamilton):** I think everything has been said, so let me add just two quick points. One is, well perhaps, we don't have many [quantum] algorithms, but that's OK, we don't have that many [quantum] computers to run them on just yet…

**<Audience laughter>**

**Fourth Pro Panelist (Alex Hamilton):** … so, you know, algorithms – it helps if we have something to do to run them with and that will probably come in time. Second thing is that, does it have to be a general-purpose quantum computer? The floating-point unit in my laptop is not general purpose. All it does is crunches numbers but it makes my games *so* much better, and I think what we really need to do is quantum gaming and that's what's really driven the microprocessor industry and that's what will drive the quantum gaming industry.

**<Audience laughter>**

**Fourth Pro Panelist (Alex Hamilton):** And finally, the classical representation of quantum computing. If you want to represent 300 qubits for a quantum computer classically, you can, but there won't be much left of the universe once you've done that.

**Chair (Charlie Doering)**: Excellent. And in time. David? Hold the mike closer.

**First Con Panelist (David Ferry):** Alright, I believe, Dan, I used the word 'practical' but there's a big difference in 'practical' and the number of the qubits you need out there. I spent a great deal of time working on quantum dots and I know how practical they are for this purpose. And there are other examples of some massively parallel analog systems, which factor the number 15 *really* fast – it's called the [human] brain.

**<Audience laughter>**

**Chair (Charlie Doering)**: Onward. Julio.

**Second Con Panelist (Julio Gea-Banacloche):** Ummm … ummm

**<Audience laughter>**

**Second Con Panelist (Julio Gea-Banacloche):** Ummm … ummm

**<Audience laughter>**

**Chair (Charlie Doering)**: Fourty seconds.

**<Audience laughter>**

**Second Con Panelist (Julio Gea-Banacloche):** Ummm … ummm

**Chair (Charlie Doering)**: Okaaaay. Now, Sergey…

**Third Con Panelist (Sergey Bezrukov):** My only point is that the solutions adopted by Nature are very, very good. For example, the other day, Laszlo Kish and I discussed the dissipation issue of 1-bit processing in our brain and in a conventional computer… and it turned out [that] our brain is 10 times more efficient in power dissipation. Why is that? For two reasons. [Firstly,] because our brain uses ten times smaller voltages. The computer uses about 1 V and the brain only about 0.1 V. The second reason is that our brain is a massively parallel computer, so that mistakes are not only prohibited but, to a degree, are welcome for our spontaneity and ability to think.

**Chair (Charlie Doering)**: Excellent. Thank you. Last one.

**Third Con Panelist (Laszlo Kish):** The brain is using noise to communicate, which is important. Concerning Hilbert space: yes classical and quantum is different. Classical is better because it is not statistical like quantum. But finally to you Charlie [Doering], your quotes are against us – I mean they are Pro! How about the moon base? In the 1970s, we expected that we would have a base on the moon at the end of the century, [which did not eventuate.] That's all.

**Chair (Charlie Doering)**: It's no good taking shots at me! I just work here.

**<Audience laughter>**

**Chair (Charlie Doering)**: What we're going to do now is … I'd like to open it up to the audience here. If people have questions, you can direct questions toward either a particular side or particular person, but we'll keep it short and then we may allow some rebuttal from the other side, whatever. So, we have a question right here.

**Audience member 1 (Unknown)**: I would like to ask a question to just anybody who's most motivated to, to one or two of you who might answer. I would like to stick to the Moon. What do you think is harder – to build a 10,000-qubit-quantum computer right now, say in the next years – some big effort – or to decide in 1960 to go and put a man on the Moon within 10 years?

**Chair (Charlie Doering)**: Who would like to take that? Carlton, it looks like you're reaching for that.

**First Pro Panelist (Carl Caves)**: No question. It's easier to put a man on the Moon. That's basically engineering. There's a huge amount of basic research that has to be done to make a quantum computer work.

**Chair (Charlie Doering)**: Anybody else? Everybody agrees. **<The whole Pro team nods affirmatively.>** That's an interesting take-home message. Derek?

**Audience member 2 (Derek Abbott, The University of Adelaide, Australia):** It seems to me, without a doubt, that small numbers of qubits have been demonstrated. So the real question for this debate should be: "Is it possible to *scale* quantum computers?" I think that's your real question and if you look at the most sensible way of scaling, which is on silicon, in my opinion – because it's a mature scaleable technology – you have then got to ask, "What is the decoherence time in silicon?" And all the papers say, "If you use pure silicon and blah, blah, blah, it's all very good." But putting my Con hat on, to help the Con team a bit…

**Chair (Charlie Doering)**: They need it.

**Audience member 2 (Derek Abbott):** … they need it, so I'm going to help them a bit. What the papers don't address is that, "OK, I've got this…"

**Chair (Charlie Doering)**: Questions cannot last longer than three minutes.

**<Audience laughter>**

**Audience member 2 (Derek Abbott):** [What the papers don't address is that,] "OK, I've got this scaleable quantum computer; I've got zillions of qubits on here; I've got all these A and J gates switching like crazy. That is a coupling into the environment. What's going to happen to that decoherence time when they are all switching like crazy? That is my question to this [Pro] side. Thank you.

**Second Pro Panelist (Daniel Lidar)**: Well, the answer once again is in quantum error correction. Provided that you can get the single qubit decoherence rate below a certain threshold, the theory of quantum error correction guarantees that you can scale-up a quantum computer.

**Fourth Pro Panelist (Alex Hamilton)**: Just to finish, to go back to your point about scalability. Although silicon is one of the things I'm working in – I don't think it's the only one that's scaleable – superconductive technology is equally scaleable. It's very good – you can go out right now and buy RSFQ <Rapid Single Flux Quantum> electronics that's basically a superconducting electronics that's been scaled-up and there's no reason that other systems can't be scaled-up. Ion traps can be put on-chip and so on. So, there's no reason that semiconductors are the only ones that are scaleable.

**Chair (Charlie Doering)**: Julio.

**Second Con Panelist (Julio Gea-Banacloche):** I think that it's always a big jump to say that just because you have demonstrated something for, say, 100 qubits that you're going to be able to scale that up 4 orders of magnitude, without encountering any unexpected problems. I don't think there are any engineers here that will support such a point of view. And, there are constraints that we can already begin to imagine, as you just mentioned. If you're going to address your qubits by frequency, for instance, it's not the same thing to have a hundred different frequencies, as it is to have a million different frequencies. And that's not all. There are constraints on the amount of energy that you need in order to perform the gates, and it's not the same to operate a hundred qubits as to operate a million of them. So, the scaling is not by any means trivial. I am willing to grant that once we have demonstrated, say, 5 qubits – with some effort in a 5-10 years time frame, we may be able to do 50 to 100 qubits.

**<Minor audience applause>**

**Chair (Charlie Doering)**: David, did you want to…?

**First Con Panelist (David Ferry):** Scaling is not all it's cracked up to be. You can go to the Intel website. You can find there a view graph, which predicts that in about 6 or 7 years from now, the power dissipation figure for your Pentium will about that of a nuclear reactor. <ftp://download.intel.com/research/silicon/TeraHertzshort.pdf, slide number 9. >

**<Audience laughter>**

**Fourth Con Panelist (Laszlo Kish):** Yeah, Alex Hamilton said in his talk yesterday, that if you use error correction you need an error rate of $10^{-6}$ or less. $10^{-6}$ *error rate* – this is a huge thing, because this is just like analog circuits, which can achieve [a] $10^{-6}$ error [rate] by using very strong negative feedback. You know, [a] $10^{-6}$ error rate [for quantum computation] seems to be hopeless. Anyway, this is very difficult.

**First Pro Panelist (Carl Caves):** I think it's generally $10^{-4}$, which is also incredibly small. But there's a lot of work in getting error correction worked out and in some systems based on dits instead of bits – that is higher dimensional quantum systems – or systems based on topological quantum computing, there's some indication that the error threshold might get up to one per cent, and then you're in the ballgame, I think. So we're just at the start of this and to dismiss the whole thing because the first results say fault tolerance is going to be extremely difficult to achieve, seems to be a mistake. Let's do some further work and see what the error threshold can get up to in other kinds of architectures and designs.

**Chair (Charlie Doering)**: OK. All right, let's move on to a different… Another question.

**Audience member 3 (Howard Wiseman, Griffith University, Australia)**: This is addressing Carl's observation comparing probabilistic computing with quantum computing. The genuine question is, "Was probabilistic computing as sexy an area as quantum computing is now?" Because it is sort of worrying that there are so many smart people working on quantum algorithms and there hasn't been another reasonable one since '97. It does indicate a genuine concern.

**Chair (Charlie Doering)**: Anyone know what the response is? David? You have the mike.

**First Con Panelist (David Ferry):** In the beginning, Popelbaum <University of Illinois> was working on probabilistic computing back in the mid-70s, around '73 or '74, and it was not a big area like this. He was kind of trudging on alone with a small, dedicated group working in it, but I don't think it grabbed the attention of big research groups around the world like quantum computing did.

**Second Pro Panelist (Daniel Lidar)**: There is a misconception that there are no good quantum algorithms out there. For problems in number theory and pure computer science, yes – there are very few. But let's not forget that quantum computers are exponentially faster at simulating quantum mechanics. Every university in the world has people in chemistry and physics departments working on trying to find fast algorithms to solve problems in quantum mechanics. A quantum computer would be enormously helpful there, so that's a huge benefit.

**Chair (Charlie Doering)**: OK.

**First Pro Panelist (Carl Caves):** I think that's a good point Howard [Wiseman]. I'm just relying on the fact that Umesh Vazirani, who has worked on both, suggested that given the current scale of effort in computer science among people who think about this, you might expect to make a big breakthrough in quantum algorithms any time or you might expect it to be in another decade. My direct response to you is that quantum mechanics is a much *richer* theory than classical probability theory, so you might think it is harder to come up with quantum algorithms, and it might take longer even with more people working on it.

**Chair (Charlie Doering)**: Another question here.

**Audience member 4 (Unknown)**: Just like to make a quick comment. Doing all those is fine but as general-purpose computers, I'm just wondering in the '60s, '70s and '80s people doing optical computing. Except for certain special purpose optical computing, there isn't any general purpose optical computing.

**Chair (Charlie Doering)**: Anybody?

**Fourth Pro Panelist (Alex Hamilton)**:  My understanding is that for optical computing… that one of the great things that it would be good for would be for Fourier transforms, and with the invention of the fast Fourier transform algorithm, there really wasn't any more need for optical computing.

**Audience member 5 (Unknown)**: That's not quite true because you've got optical parallelism and your Fourier transform [is traditionally computed in] series.

**Audience member 6 (Kotik Lee, BAH, USA)**: Optical computers are used extensively with defense systems, for special purpose processes.

**Chair (Charlie Doering)**: Another question. Up the back.

**Audience member 7 (Fred Green, University of New South Wales, Australia)**: Well, it's really a comment. It's another take on the relative lack of algorithms. One of the things that happens when you make a machine that's enormously complex is, that it may well become something that uses its emergent behaviour – to copy a buzzword. The thing is, that in a sense we are thinking in a reductionist way about machines, we're thinking of a specification and rules and designs to make an enormous machine, but it's equally likely that the machine will go and do things that you simply cannot predict from its underlying equations. That is just an open question. For example, you cannot – just by having a set of equations and putting it on a computer – you cannot get superconductivity out of that. Something has to make it all complete and all I'm doing is actually repeating what Laughlin said some years ago now. It's quite conceivable that a machine in all its complexity will be able to do things like that. It's something that human brains are quite good at.

**Chair (Charlie Doering)**: Any response? Julio?

**Second Con Panelist (Julio Gea-Banacloche):** I'll venture a response. That's certainly a possibility but it's not currently an envisioned possibility, the way people envision this huge fault-tolerant quantum computer. Most of its time – 99.99% of its time in every clock cycle it will not be doing anything except error correction. Emergent behaviour would be, you know, remarkable – almost anything could show emergent behavior more likely than such a machine.

**Chair (Charlie Doering)**: You Carl, you have to respond to that.

**First Pro Panelist (Carl Caves):** I'm not really directly responding to that. I want to say something that popped into my head that has something to do with that. Now what if there were a fundamental decoherence mechanism in the universe that couldn't be explained by coupling to external systems. You could error correct that. Wouldn't that be pretty neat? You could restore linear quantum mechanics even though the universe is fundamentally not linear quantum mechanics.

**<Audience laughter>**

**Chair (Charlie Doering)**: Whoa, whoa, whoa. OK.

**Second Pro Panelist (Daniel Lidar):** I just wanted to say something about the [observation that] 99% of the time is spent doing error correction. This is true, but it does in no way contradict the fact that a quantum computer offers a speed up.

**Audience member 8 (Michael Weissman, Univ. Illinois Urbana Champagne, USA)**: I did not understand those last remarks [of Caves on restoring linearity], but while we are on the same topic: you mentioned earlier that the interpretation of quantum mechanics does not affect the operation of a quantum computer. That is certainly true up to an extent. However, if there were an intrinsic non-unitary operator involved, at some point, for example, you would find decoherence in cases where [you] do not expect decoherence if there are only unitary operators. If you made a quantum computer more or less of the same physical scale as your head and with similar amounts of mass and current involved in its thoughts as in yours and it did not show unknown non-unitary operations that would be very important for understanding quantum mechanics.

If it did, it would be even more important because it would support the idea that some modification is needed. Either way, conceivably, it would have something to do with the experimental realisation of tests of modification-type interpretations.

**First Pro Panelist (Carl Caves):** Yeah, I certainly agree with you that one thing you might find out is that there are fundamental non-unitary processes when you get a sufficiently large system, and those are processes responsible for making the world classical and they would represent a barrier to making a quantum computer of sufficient size. Those are important issues. I don't call them interpretational because they're changing quantum mechanics, whereas when I refer to the interpretation of quantum mechanics I mean keeping what we've got and figuring out what it means – not making changes.

**Chair (Charlie Doering)**: As the Chair of this session, I'm going to declare we're going to keep quantum mechanics the same way, as it's not fair to try to change quantum mechanics for either the Pro side or the Con side.

**<Audience laughter>**

**Chair (Charlie Doering):** OK, yes, absolutely, absolutely.

**Third Pro Panelist (Howard Brandt):** We've got some affirmation of the worth of the pursuit of quantum information science and technology by one of the editors of *Physical Review A*, Julio here. But Charles, one of the things I hear is you're an editor of *Physics Letters A*. I'm very concerned about one of the things you said previously and that the editor of *Physics Letters A* may be ignorant about quantum mechanics. You stated that yourself!

**<Audience laughter>**

**Chair (Charlie Doering)**: Very good. I apologize.

**<Audience laughter>**

**Chair (Charlie Doering)**: The great thing about *Physics Letters A* is that each editor has their own area of expertise and mine is explicitly not quantum information, [which means I have no bias]. So, anyway, quantum mechanics shall not be changed in remaining discussion and we'll move on from there. Peter Hänggi.

**Audience member 9 (Peter Hänggi, University of Augsburg, Germany)**: I like quantum computing because you can see all this knowledge being brought together from different areas of science and great progress in understanding quantum mechanics. But I also believe in human nature, you know. After five, ten years I get a bit tired because I've seen enough of it. See, there are those things that can be done quickly, and I'm not so sure this momentum carries on when it comes to do the very hard work. Most of the people here don't want to do the nitty-gritty work, the core and the details about this stuff, and so on. I think the excitement, which is so high up with tackling all these problems on quantum information, will eventually slow down. [If] the problems are not [solved in] three, four, five years maximum, and then of course we need something else and we don't know what the next excitement in science will be; but most likely we physicists don't want to do for two or even more years [the] nitty-gritty work on a detail. Moreover, we also need to talk about the engineering, so that explaining this heightened expectation [*sic*]; we also need practical things from this whole exciting quantum computer and computation [area].

**Chair (Charlie Doering)**: OK, so that sounded like a … ? Is that a… ? Could we have a response to that? Is that a Pro or a Con?

**Audience member 9 (Peter Hänggi)**: I don't know what it is.

**Chair (Charlie Doering)**: Yeah, that's OK.

**<Audience laughter>**

**Second Pro Panelist (Daniel Lidar)**: I think it would be great [if people got tired,] because there are way too many papers in this field right now.

**<Audience laughter>**

**Chair (Charlie Doering)**: <Julio nods affirmatively. > OK, the editor of *Physical Rev. A* seconds the motion.

**<Audience laughter>**

**First Pro Panelist (Carl Caves)**: I think the example of quantum cryptography shows that people are willing to do very sophisticated, higher mathematical and physical work on a system that is closer to the point of transference into something useful. And I think that's a good example of quantum cryptography inspiring extremely useful work – detailed work about improving security in quantum crypto systems – for real systems. So I think that as long as the experimental work in the field is moving forward to increasing numbers of qubits, there are going to be important theoretical problems to address, and we have plenty of theorists to work on them, and I think they will.

**Chair (Charlie Doering)**: Yup.

**Second Con Panelist (Julio Gea-Banacloche)**: I think that the concern is more with the funding agencies losing interest and… clearly, if this machine is 20 or 30 years in the future, I think that it doesn't take a prophet to predict, that they are not going to continue the level of funding… the current level of funding for the next 20 or 30 years. Moreover, as I said, without any more algorithms there is the possibility that they will lose interest much earlier because all we need is basically an easy, convenient alternative to RSA encryption and you're in business, and there are already encryption – public key encryption – algorithms that nobody knows whether they are equivalent to factoring or not. Which means that even if you had the big quantum computer *today*, you would not know how to use it to crack those forms of encryption. So, it's really only a matter of time. So…

**Chair (Charlie Doering)**: Howard? Comment?

**Third Pro Panelist (Howard Brandt)**: [Regarding] the business about, you know, if it's going to take thirty years to build a quantum computer, that the government agencies aren't going to wait that long and continue to fund it. I don't believe that, because the imperative is still considerable. Witness thermonuclear fusion. Sakharov came up with the invention of the Tokomak. Now, that was a long, long time ago. That has continued to be funded. Also, newly, and nicely, inertial confinement fusion. And you know, I remember in the '70s I was asked to predict when we would have controlled thermonuclear fusion, and I said, 'At the earliest 2030', and certainly after all the major participants are *dead*. And that is true of large-scale quantum computers too. The government will still have this imperative and it will be supported at some level, I believe.

**Chair (Charlie Doering)**: Now let Julio talk.

**Second Con Panelist (Julio Gea-Banacloche):** Now I think there's a big difference between physical controlled fusion and quantum computing, as we know it now. I mean, once you get a fusion reactor going, then you can do a lot of things with the energy. Once you get these huge quantum computers going you can do exactly one thing …

**<Audience laughter>**

**Second Con Panelist (Julio Gea-Banacloche):** No, sorry—apologies to Daniel, actually—*one* thing of *strategic importance*, OK? Which is to break the RSA code. How much longer is RSA encryption going to be of strategic importance? My guess is not 30 years, OK. Now, I completely agree with Daniel, this is of extremely high [scientific value] and I hope the NSF will continue to support the development of quantum computers at the medium-sized scale for all the universities that will want to have a quantum computer.

**Chair (Charlie Doering)**: Alex?

**Fourth Pro Panelist (Alex Hamilton):** Well, I think my point has been said, actually. It's not just the one algorithm you want to include – there's a whole raft of fundamental science reasons, there's a whole raft of computational reasons that you wanted as well for simulating physical systems. I mean, it's crazy that we have a transistor with 50 electrons in it and we still can't calculate, properly, what its properties should be from a fully quantum mechanical viewpoint. So that would be kind of nice, and … the second thing is about, going back to RSA, if everyone switched to quantum hard codes there'd be no need for this computer, but wouldn't you *love* to know what Clinton *really* said about Lewinsky? I mean, you could go back and …

**<Audience laughter>**

**Chair (Charlie Doering)**: OK, OK, let's keep this clean!

**<Audience laughter>**

**Chair (Charlie Doering)**: Let's move on here. Anybody have a comment or complaint, or a … you know. Gottfried.

**Audience member 10 (Gottfried Mayer-Kress, Penn State University, USA)**: Yes, just a question or comment on the statement about the brain and it was also very … so sweeping a rejection of any kind of possibility of quantum computation in the brain, and you gave the impression [that] everything was known how the brain works and, you know, there's really no open questions, so do you really know how we make a decision? How you make a choice between different alternatives and the speed at which this is happening? So, it seems to me like … just from the problem solving point of view: if you think about it, how fast a human brain, can select from a huge database of visual or sensory inputs and make a very rapid decision. I mean, that sounds very much like a quantum computation to me, and if you go down to the biochemical processes of how ion-channels open and close I think, you know, that quantum processes certainly play a role. So, I don't understand why you just completely reject the possibility of quantum computation occurring.

**Third Con Panelist (Sergei Bezrukov):** I agree with you that we don't understand how our brain operates in… concerning what you just said. My only point is that according to the current knowledge of the 'elemental base' of the brain, responsible for logical operations, there is no place for quantum computing.

**Chair (Charlie Doering)**: Carlton?

**First Pro Panelist (Carl Caves):** I used to have the conventional view – and I still have it – that the probability is about [point] 50 nines in a row that there aren't any coherent quantum processes going on in the brain that are of any value. You can do simple calculations that show that decoherence removes any coherent quantum information processes in the brain. But we now know that in complex quantum systems there are these decoherence free subspaces just sitting around that are free of certain kinds of decoherence and it's not out of the question that maybe something's going on there and, you know, evolution by natural selection is awfully good at figuring out how to do stuff. I'll give it a probability of epsilon, where epsilon is smaller than the error threshold, but I wouldn't rule it out.

**Chair (Charlie Doering)**: Interesting point. Anybody else? Yeah, let me see your hand.

**Audience member 11 (Howard Wiseman, Griffith University, Australia)**: I'll keep supporting the Con side, just to be fair. Daniel, you keep bringing up this simulating quantum systems thing, but how big … given that classical computers will probably keep going faster for the next, say, 20 years … how big a quantum computer do you actually need to make it useful? To make it definitely useful? And, you know, is there anything that can bridge the gap between, you know, the next 5 to 10 years, and that sort of level?

**Second Pro Panelist (Daniel Lidar):** Well, there are several papers, which have looked at this question in detail, and not taking into account the error correction overhead, it turns out that at about 100 qubits you can solve problems in mesoscopic quantum physics, which are not possible on any reasonable classical computer. So, 100 qubits is my answer but you'd have to multiply that probably by a factor of like at least 15 if you want to take error correction into account.

**Audience member 11 (Howard Wiseman)**:  Is there something that can take us from where we will be in the foreseeable future to that level of some 1500 qubits?

**Chair (Charlie Doering)**: Did you hear the question?

**Audience member 11 (Howard Wiseman)**:  What is going to motivate us to go from the level of having 10 or 100 qubits – where we can do interesting things from the point of communications and distillation and stuff like that – to that level, which is considerably harder?

**Second Pro Panelist (Dan Lidar):** Well, one problem, for example, is understanding superconductivity in metallic grains. So, if you… if that is a problem that is of considerable interest, which I believe it probably is, I can see that motivating going to that number of qubits that's required, and there are plenty of other problems in this class of highly-correlated-electron systems that are mesoscopic, for which you would need a quantum computer on the order of 100-1000 qubits.

**Chair (Charlie Doering):** Another question here.

**First Pro Panelist (Carl Caves):** Can I say one more thing about that?

**Chair (Charlie Doering):** Certainly.

**First Pro Panelist (Carl Caves):** Let me say something pretty quickly. The systems that are proposed for quantum computing and quantum information processing are the cleanest we know of. The best records for quantum coherence are the atomic physics systems, now using trapped ions and trapped neutrals. Those are pristine systems for which decoherence is very low, but it's not so clear how you scale those. The condensed systems are easier to see how to scale because they rely on more conventional technology, but their record for decoherence isn't as good. All the superconducting qubits are getting there now in terms of decoherence, but we still have to see how they do when they're coupled together. We don't yet know which one of these systems, if any, is going to be one that ultimately works out. We don't know what the architecture is going to look like for a 1500 qubit quantum computer.

**Audience member 12 (Unknown):** Yes, I would like to ask Carlton a question. You seem to be hopping around a bit together with other members in the Pro team, appeasing the two editors <of PRA and PLA>. One [statement] is that in principle you can do [quantum computing] operations with error correction… Simply, it sounds like it's just an engineering problem; get enough engineers together and enough money together and it [quantum error correction] will work. And on the other hand the suggestion is that you actually need some more basic research – you may find that, you know, you run into something like a mental limitation. So like, the question is: Is it just an engineering problem or not?

**First Pro Panelist (Carl Caves):** You might have exposed a rift in the Pro team, I don't know. We're sitting awfully close together up here, so if you can see any rift between us … but I think there's a lot of basic research to be done. It's not an engineering problem yet. I think a fairer comparison, when I was asked about the space program, would have been the Manhattan Project. If you put in an amount of money comparable to that in today's dollars, would we get a quantum computer a lot faster? Which would be several billion dollars a year, I reckon. Oh, I don't think so. I think we wouldn't know what to do with it, because it's not yet an engineering problem. There's a lot of basic science yet to be done before we know which physical system is the best one.

**Chair (Charlie Doering):** So, we have something from the Pro side?

**First Pro Panelist (Carl Caves):** Yeah.

**<Audience laughter>**

**Second Pro Panelist (Dan Lidar):** Alright, I think you are probably referring to … or extrapolating from a comment that I made that, well, we have error correction, therefore problem solved. No, that's not the case. The fact that we have an existence proof or a viability proof, if you wish, that quantum computers are possible, does not in any sense imply that there's no basic research left to be done. I mean, it's like – what's a good analog? – maybe: an existence proof is like saying that we have an axiomatic system for doing mathematics and now, that's it, we're done. Of course, a lot of theorems remain to be discovered. There's a lot of basic research to be done on how we can actually construct a device and there'll also be lots of spin-offs in terms of just interesting fundamental questions that are not necessarily related to how you construct a device.

**Third Pro Panelist (Howard Brandt):** I agree with Carl. I think Derek's comment was very appropriate and it sort of addresses this question of the fundamental nature of current research. [Regarding] Derek's comment, well you know, the hybrid Kane-type quantum computer in silicon, and other solid-state approaches that we include here, like quantum dots, well, they're scalable. What does that mean in practice? It means that there's a giga-dollar industry in semiconductors and solid-state, and frankly, you know, I think that the funding agencies sort of translate that into scalability. I mean, after all, you know, the classical widgets scaled, so we make one quantum widget and put the widgets together, but as Derek sensibly questioned, you know, right now, be it Josephson junctions, quantum dots or a Kane-type of quantum computer, the study of decoherence is at a very primitive level. The study of how to produce controlled entanglement is at a very primitive level. Gates and solid-state approaches are at a very, very primitive level. People do not know how to do this. They're doing research to hopefully, you know, be able to do this, but it's a big question mark. It's a basic research issue. And so Derek, you know, is justified in questioning the scalability, of the solid-state semiconductor approaches anyway. So, it's a basic research issue. The answers are not there. If they were, we'd hear about it in program reviews. I mean, I've heard tens and tens of program reviews, and nobody is coming close yet, but that doesn't mean they won't. Basic research is needed in fact.

**Chair (Charlie Doering):** Any other comments or questions from the audience? Derek wants to rebuff.

**Audience member 13 (Derek Abbott, The University of Adelaide, Australia):** [No, but] I just thought I'd give the Con team some more help because they need it.

**<Audience laughter>**

**Second Con Panelist (Julio Gea-Banacloche):** I wish you would stop saying that.

**<Audience laughter>**

**Audience member 13 (Derek Abbott):** OK, so I think we've established that the real question is: scaling. To make it practical we need a scaleable quantum computer. To make it scaleable you're talking about chip technology for a number of reasons because it's the only way we know how to make scaleable things – and we've got millions of dollars of backing behind that. Now, as soon as you put qubits on a chip and line them up in a nice little pretty order, I find it very hard to believe that you can make a useful quantum computer with that, because on top of those qubits you're going to need classical control registers to control the gates and you are going to need read-out circuitry. So there's going to a number of post-processing steps on top of that. Now, I know some clever guys have put phosphorous ions [on chip] nicely in a neat little row and it works. But once you do all that post-processing, are they [the qubits] really going to stay still? So, this is my question to the Pro team.

**Chair (Charlie Doering)**: Yes, and it's a good question. Alex?

**Fourth Pro Panelist (Alex Hamilton):** OK, so for the specific case of phosphorous and silicon it actually looks like they do stay put, during the post-processing – but that's just a specific answer. But the more general answer is, you do need control chip circuitry, absolutely, and … so there's no reason that that has to be on the same chip. There's no reason that we can't build a [separate] complete high-frequency silicon germanium control electronics [chip] and match it up.

**Audience member 13 (Derek Abbott):** You have to use the same chip because of noise.

**Fourth Pro Panelist (Alex Hamilton):** No, no, it doesn't have to be on the same chip, Derek, because they [only] have to be physically close to each other. They don't have to be *on* the same chip.

**Chair (Charlie Doering)**: OK. He says no and you say yes.

**<Audience laughter>**

**Second Con Panelist (Julio Gea-Banacloche):** Yes, I wanted to say something about that too, regarding the same sort of thing, the control systems. I actually gave a talk yesterday on this subject and there are constraints there: how large the control systems have to be in some sense. So in some sense, some of these amusing quotes **<referring to the famous quotes recalled by Charlie Doering>** are a little misleading because …

**Chair (Charlie Doering)**: We don't know.

**Second Con Panelist (Julio Gea-Banacloche):** …they suggest that, you know, quantum computing might follow a path similar to classical computers where you start with something huge, like vacuum tubes, and then slowly and over time, sometimes fast, you start making things smaller and more efficient and so forth. And the indications are that it's not going to be like that. I mean, when we get the quantum computer we're stuck at the vacuum tube level. The control systems have to be large because they have to be classical, so there is going to be no pocket quantum computer that somebody will be able to carry around, and there are minimum energy requirements and so another question to ask, you know, is: how are you going to deal with the heating and so on? How are you going to extract it?

**Chair (Charlie Doering)**: OK. Laszlo's going to make one more comment and then we're going to move into the next stage, the final stage of this panel.

**Fourth Con Panelist (Laszlo Kish):** [Regarding] the comment of Julio's… the calculations I showed yesterday – at the same temperature, same speed and same number of elements – the quantum computer dissipates more energy when processing the same information. So, again, the dissipation and noise is the key [as to why classical computers are better].

**Chair (Charlie Doering)**: What I would like to….you're going to talk? I think we need to move on …

**First Pro Panelist (Carl Caves):** Well, we're at about the last stage …

**Chair (Charlie Doering)**: OK.

**First Pro Panelist (Carl Caves):** We might have to run the gauntlet to get out of here.

**Chair (Charlie Doering)**: OK. That's right. Let me organise the last stage as follows. Let me have each one talking. Think about a *two*-sentence summary of your view, a two-sentence summary of your view, and we'll run down the road here and then we'll take a vote on how the audience feels what the panel says. A forum – OK? Carl.

**First Pro Panelist (Carl Caves):** Well, my consistent view has been that it would be extremely difficult to build a general-purpose quantum computer though it might be somewhat easier to build quantum simulators, but that's not the point of why information science – quantum information science – is a discipline worth pursuing.

**Chair (Charlie Doering)**: That's one sentence. OK, good.

**<Audience laughter>**

**Chair (Charlie Doering)**: Daniel.

**Second Pro Panelist (Dan Lidar):** I agree.

**Third Pro Panelist (Howard Brandt):** Well, again, no one has demonstrated that a large-scale quantum computer is, you know, physically impossible, and certainly small-scale quantum information processors are possible and have already been demonstrated. It's a worthwhile enterprise and will continue.

**Fourth Pro Panelist (Alex Hamilton):** We're scientists. Our job is to try and understand nature and if we want to – and we're humans – and we try to control nature. And we've been given this amazing curiosity and we want to do things. Why does one climb Mount Everest? Because it's there. **<The Con team all purse their lips in a thoughtful pose, gently nodding, apparently conceding this point. >** If we build this thing [a quantum computer], let's have a go or let's prove that it's simply not physically possible.

**Chair (Charlie Doering)**: Very good. Julio?

**Second Con Panelist (Julio Gea-Banacloche):** I really would like to say that I am also very impressed …

**Chair (Charlie Doering)**: Julio, also let me just say, that the Con team does not have the volume of the Pro team….

**Julio**: OK.

**Chair (Charlie Doering)**: …we would like to hear how the argument's going. Hold the mike closer!

**Second Con Panelist (Julio Gea-Banacloche):** Well, I would really like to say that I find it incredibly amazing and very, very impressive. [A lot of] good science has come out of quantum information, for the past seven years. And if quantum computing is responsible for this, then it's a good thing. The dream, at least, [is] of a quantum computer.

**Third Con Panelist (Sergey Bezrukov):** While [many] functional processes in the brain are *not* understood, the 'elemental base' [of the brain] is very well studied. Also, main interactions between the elements are already well understood. So, as I said, there is no any [*sic*] place for quantum computing in the human brain. But, the concepts, which are being developed by the scientists working in this field, will find their way into the brain studies and will be very useful there.

**Chair (Charlie Doering)**: Yes.

**Fourth Con Panelist (Laszlo Kish):** Quantum compared to classical. Quantum means more noise, statistical in nature, more dissipation and higher price.

**<Audience laughter>**

**Chair (Charlie Doering)**: OK, that was enough. Now we've got first … for the record … I guess the question is a 'pro'-'con'/'dream'-'reality' thing. I would like to take a show of hands for … first, question number one. How many people think that quantum computing is really a dream and it's just going to fall by the wayside and our attention will go some place else, and … Can I have a show of hands?

**<A few hands show>**

**Chair (Charlie Doering)**: OK, how many people think that it's possible that quantum computers as – people envision it as a *tool* - is simply not going to happen? The way we're visioning it now?

**<A few hands show, some people holding up two hands>**

**Chair (Charlie Doering)**: You are not allowed to hold up two hands. But you can attempt to have your hands in a superposition of "for" and "against."

**<Audience laughter>**

**Chair (Charlie Doering)**: That's good, that's good. I'm impressed! Now, now … but that does not mean that the complement to that set will be reality, OK. So, how many people think that there's a possibility that it may be a useful tool, based on the ideas that we're now tossing around in the year 2003? That it's going to emerge – and some of us in this room are [still] alive to realise it? Anybody agree with that?

**<A unanimous majority of hands show>**

**Chair (Charlie Doering)**: OK. I think the conclusion is clear. I would just like to reinforce the whole idea of predicting the future in computer science [is dangerous]:

"640 K ought to be enough for anybody." – Bill Gates, 1981.

Let's thank everyone on the panel here.

**<Audience applause>**

**<End of transcript>.**

**<u>Acknowledgements</u>**

The assistance of Phil Thomas, as the audio transcript typist and amanuensis, and Gottfried Mayer-Kress, for the mp3 recording, are gratefully acknowledged. The mp3 recording can be downloaded from the *Complexity Digest* at: http://www.comdig2.de/Conf/SPIE2003/

Thanks are due to the many people who proof read this transcript; any remaining errors are of course mine (Derek Abbott).